\documentclass[preprint,3p,times,12pt]{elsarticle}

\usepackage{verbatim}
\usepackage{graphicx}

\usepackage{float}

\usepackage{bm}
\usepackage{amsmath}
\usepackage{amssymb}
\usepackage{epsfig}
\usepackage{multirow}

\def\bV{{\bf{V}}}
\def\bA{{\bf{A}}}
\def\bS{{\bf{S}}}
\def\bX{{\bf{X}}}
\def\bY{{\bf{Y}}}
\def\bB{{\bf{B}}}

\def\bu{{\bf{u}}}
\def\bv{{\bf{v}}}

\def\bT{{\bf{\Theta}}}
\def\bF{{\bf{\Phi}}}

\def\be{\begin{equation}}
\def\en{\end{equation}}

\DeclareMathOperator{\E}{\mathbb{E}}
\DeclareMathOperator*{\argmin}{argmin} 

\def\RR{\rm \hbox{I\kern-.2em\hbox{R}}}
\def\NN{\rm \hbox{I\kern-.2em\hbox{N}}}
\def\ZZ{\rm {{Z}\kern-.28em{Z}}}
\def\CC{\rm \hbox{C\kern -.5em {\raise .32ex \hbox{$\scriptscriptstyle
				|$}}\kern
		-.22em{\raise .6ex \hbox{$\scriptscriptstyle |$}}\kern .4em}}

\def\<{\langle}
\def\>{\rangle}

\long\def\symbolfootnote[#1]#2{\begingroup%
	\def\thefootnote{\fnsymbol{footnote}}\footnote[#1]{#2}\endgroup}

\begin{document}

\begin{frontmatter}
	
\title{Leveraging data from nearby stations to improve short-term wind speed forecasts}

\author[]{R. Ba\"{\i}le\corref{cor1}}
\ead{baile@univ-corse.fr}
\author[]{J.F. Muzy}
\ead{muzy@univ-corse.fr}

\address{\footnotesize \em Laboratoire ``Sciences Pour l'Environnement'' \\	\footnotesize \em UMR 6134 CNRS - University of Corsica \\
	\footnotesize \em Campus Grimaldi, 20250 Corte (France)}

\cortext[cor1]{Corresponding author}


	\begin{abstract}
    In this paper, we address the issue of short-term wind speed 
    prediction at a given site. We show that, when one uses spatiotemporal information as provided by wind data of neighboring stations, one significantly improves the prediction quality. Our methodology does not focus on any peculiar forecasting model but rather considers a set of various prediction methods, from a very basic linear regression to different machine learning models. In each case, our approach consists in specifically and incrementally studying the benefits of using wind data of the surrounding stations. We show that, at all horizons ranging from 1 to 6 hours ahead, the relative gain on the RMSE of the predicted wind speed can increase up to 20 \%. For all the considered forecasting methods, we show that such a gain is far better than the one obtained by considering other kind of information like local weather variables or seeking for an optimal deep learning model. Moreover we provide evidence that non-linear models, as neural networks or gradient boosting methods, significantly outperform linear regression. These conclusions are simply interpreted as resulting from the ability of a method to capture the transport of the information by the main flow in the upwind direction.

	\end{abstract}

\end{frontmatter}

\section{Introduction}

Besides all societal and economic issues related to weather prediction, wind speed forecasting is particularly important for the needs of energy production. Indeed, renewable energy and, notably wind power, represents an increasingly large part of the global electrical power generation. Such a process, however, strongly depends on the volatility of the resource. Indeed, the stochastic and highly intermittent nature of surface wind velocity fluctuations leads to a great amount of uncertainty in wind energy production. In that respect, to achieve large-scale integration of wind energy in a power grid, accurate wind speed predictions and thereby forecasts of the power output of wind farms, are a key element for the energy suppliers since this may strongly affect the decision-making processes (economic dispatch, reserve allocation or power exchanges with neighboring system). The challenging task of designing efficient tools that provide accurate wind speed forecasts has therefore motivated, during the past two decades, a very large number of studies (see, e.g., \cite{7764085,hashli20,5619586,GIEBEL201759,kapisigi04}). Usually, prediction methods are categorized into two main classes. First, ``Numerical Weather Prediction" (NWP) methods mainly consist in integrating, from a given initial state, a large system of non-linear equations deduced from a specific model describing the physical and chemical processes governing the atmosphere/ocean dynamics \cite{NWPBook}. Despite recent advances in that field \cite{bauer15}, NWPs require a very large amount of computational resources and are thus only suited for medium to long term predictions (i.e. from a few hours to a few days). As far as short-term predictions are concerned (from few minutes to few hours), one usually prefers ``statistical'' or ``data-driven'' approaches. These methods mainly rely on some model or statistical inference method that uses past data to provide a forecast of future observations. Such approaches can include, for instance, the design of peculiar stochastic models for the observed fluctuations, like those related to the domain of time series analysis that can be calibrated on historical data (see e.g., \cite{poggi,MuBaPo10,BaMuPo11,arimaWind16}). 
Alternatively, many approaches rely on ``Machine Learning'' (ML) methods. During the past decade, we have witnessed the explosion of the usage of ML techniques and especially Deep Neural Networks (DNN) or Deep Learning (DL) in a wide range of areas and notably for short-term wind speed prediction. Many neural network architectures have been proposed including classical Recurrent Neural Networks \cite{lstm_ws_2018,lstm_2021}, Temporal Convolutional Neural Networks \cite{tcn_2020,mcn_ws_2020} or Graph Convolution Networks \cite{gcn_ws_2018,gcn_ws_2019,gcn_ws_2021} just to mention a few very recent examples. We refer the reader to Refs. \cite{hashli20,Manero18,Manero19} for an overview of this ongoing, very active topic.

In this article, our main goal is to show that using information about the spatiotemporal wind distribution as provided by the observations coming from various stations neighboring a given site, allows one to greatly improve the short-term prediction at this site. The idea to leverage Space-time information to improve the prediction at a given location has been explored by a few recent studies but they are all focused on a particular model and its comparison against pre-existing simple approaches.  
In \cite{Gneiting06}, the authors propose a ``regime-switching" space-time model to obtain accurate and calibrated, fully probabilistic forecasts of wind speed or wind power. Their model is among the first ones that accounts for temporal and spatial correlations of wind velocities: geographically dispersed meteorological observations in the vicinity 
of the wind farm are used as off-site predictors. This approach is further developed and improved in \cite{Hering2010}. In \cite{Hill2011}, the authors also account for the correlation of wind speed between different geographical areas by proposing a vector auto-regressive model (VAR). They show that such a model allows one to improve short-term forecasting as respect to mono-variate approaches, as illustrated by wind speed data dispersed over the United Kingdom. In \cite{TastuPinson14}, the authors provide a probabilistic wind power forecasting for a single site of interest while using information from other wind farms as explanatory variables. Their methodology to construct the predictive densities can be either parametric or non-parametric but in both cases, they show 
on their test set of wind farms in Denmark that accounting for spatiotemporal effects 
improves the quality of probabilistic forecasts for
a range of lead times up to several hours. A spatiotemporal predictor of wind speed relying on the estimation of a wind-regime dependent covariance matrix of the complex vector (representing wind amplitude and direction) at various sites 
is proposed in \cite{Browell14} and is found to considerably improve simpler methods.
In \cite{Filik16}, another linear model (``multichannel ARMA'') is introduced to account for the temporal cross-dependencies of wind speed at a target location and a set of neighboring measurements. This model is tested using real wind data collected at five stations in Turkey and is shown to provide considerable improvements over well-known standard approaches.
In Ref. \cite{ZhuQiCh18}, the authors show that when using the full set of data from a regular array of wind turbines (as e.g. distributed over a given wind farm grid) through a Neural Network  designed to capture both spatial and temporal dependencies (mainly a CNN coupled with a fully connected layer), one improves the wind speed prediction at very short range (few minutes to one hour). In \cite{gcn_ws_2018}, the authors propose a deep learning framework that accounts for temporal dependencies through a LSTM while it captures the spatial dependencies between various locations using a Graph Convolution Neural Network. They show that their method is very efficient on weather data and notably for wind speed prediction on GSOD dataset. In \cite{gcn_ws_2019,gcn_ws_2021} similar approaches based on Graph Convolution Network are proposed in order to improve wind speed predictions and in both papers, it appears that capturing spatiotemporal features from a set of stations leads to better forecasting performances. In Ref. \cite{Zheng21}, the authors propose a prediction method based on the so-called ``capsule network'' originally introduced for image or movie processing, for geographically dispersed wind farms over a region.  Using wind data from multiple wind farms in Ohio, the authors demonstrated that their approach outperforms previous forecasting methods. Let us finally cite Ref. \cite{Lars2006} that is one of the pioneering works advocating the use of geographically dispersed meteorological observations at upwind sites to improve short-range forecasts but is also the single one that focuses
more on data than on a specific model. The authors investigate the use of ``off-site observations'' as predictors in statistical forecast techniques like linear regression,
Support Vector Machines or Feed Forward Neural networks. However their study relies only on 3 distant sites over few months and only address 2-hours forecasting horizon.

We see in the previous literature review that studies that consider the possibility of accounting for neighboring observations in order to improve wind speed forecasting, are mostly focused on promoting a particular prevision model which is compared with standard approaches.
None of them (with one exception) is mainly centered on data with the goal of pointing out the interest of using spatiotemporal information, {\em independently of the considered model}. This is precisely our main goal in this paper. Our ambition is not to propose yet another prediction method that would outperform all the existing ones but rather study
in what respect taking into account the wind data of neighboring locations can improve the quality of short-term prediction. For that purpose, we consider a set of various forecasting methods that we also want to compare to each other and to the reference  ``persistence'' method. We aim at estimating the improvement of adding incrementally such a spatial information. The so-obtained improvement will be notably compared to the one obtained by using complementary weather variables such as temperature, pressure or humidity. We also want to discuss, by analyzing the performances of various Neural Network models, the potential improvement of choosing one particular architecture as compared to the benefits obtained by exploiting all available data from surrounding stations. Our second goal is to show that ``non-linear" methods such as those represented by Deep-Neural Networks are much more efficient than linear regressions in the sense that the former can better capture the wind transport which is essentially wind regime dependent and thus a non-linear phenomenon.

The paper is structured as follows: In Section 2, we describe the data set we use within this study while Section 3 is devoted to the description of various machine learning methods and notably the deep neural network models we consider. In Section 4, we present the main results of the application of these forecasting models over the set of 9 representative stations. 
The conclusion and some prospects for future research are provided in Section 5.

\section{The KNMI data set}
\label{sec:data}

\begin{figure}[t]
	\centering
	\hspace*{-0.4cm}
	\includegraphics[height=9cm,width=10cm,angle=0]{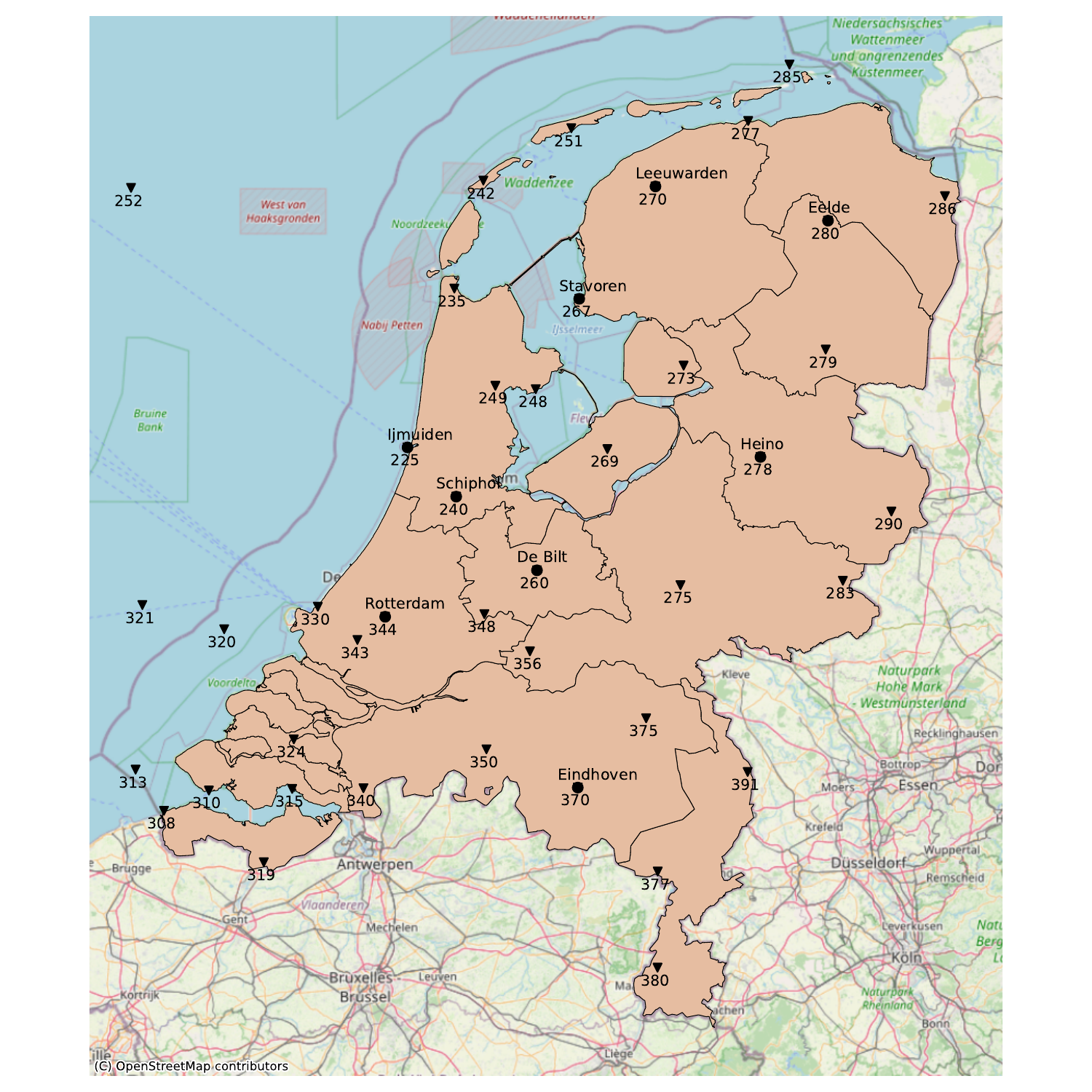}
	
	\caption{The spatial distribution of the 42 stations over the Netherlands we use from the freely available Royal Netherlands Meteorological Institute wind series database. The symbols ($\bullet$) represent the 9 ``reference'' stations where wind speed prediction is performed while symbols ($\blacktriangledown$) represent the other nearby stations whose data are used as model input.}
	\label{fig_carte}
\end{figure}

\begin{table}[h]
	\centering
	\resizebox{\columnwidth}{!}{%
		\begin{tabular}{|l||l|l|l|l|l|l|l|l|l|}
			\hline 
			\multicolumn{1}{|c||}{Station}         & \multicolumn{1}{|c|}{225} &
			\multicolumn{1}{|c|}{240} & \multicolumn{1}{|c|}{260}        & \multicolumn{1}{|c|}{267}         &
			\multicolumn{1}{|c|}{270}  & \multicolumn{1}{|c|}{278}   & \multicolumn{1}{|c|}{280} & \multicolumn{1}{|c|}{344}  & \multicolumn{1}{|c|}{370}\\ \hline 
			Wind mast height (m) & 18.5 & 10 & 20 & 10 & 6 & 10 & 10 & 10  & 10 \\ \hline
			$\overline{V}$ (m/s) & 7.26 & 4.89& 3.37 & 5.62& 4.64 & 3.04 & 4.11 & 4.31  & 3.72 \\ \hline
			$V_{\max}$ (m/s) & 28 & 23 & 15 & 24  & 21 & 18 & 20 & 21  & 18 \\ \hline
			$N_{\mbox{s}}$ (training) & 157,776 & 157,776 & 157,776  & 157,776 & 157,776 & 157,776 & 157,776 & 157,776  & 157,775 \\ \hline
			$N_{\mbox{s}}$ (validation)  &  26,304 &  26,304 &   26,304   &  26,304   &  26,304 &  26,304   & 26,304  & 26,304  & 26,304    \\ \hline 
		\end{tabular}
	}  
	\vspace*{-0.1cm}
	\caption{Main statistical features of the 9 reference stations. In each case, wind amplitude (hourly mean) and direction (last $10$-min mean speed of each hourly period) are available. These data are available (with no missing value) over the whole period 2001-2021 which corresponds to a total of 184080 hourly periods, 157776 for the model training sample and 26304 for the validation sample)}
	\label{tab:dataset}
\end{table}

\begin{figure}[h]
	\centering
	\hspace*{-0.4cm}
	\includegraphics[height = 12cm,width=14cm,angle=0]{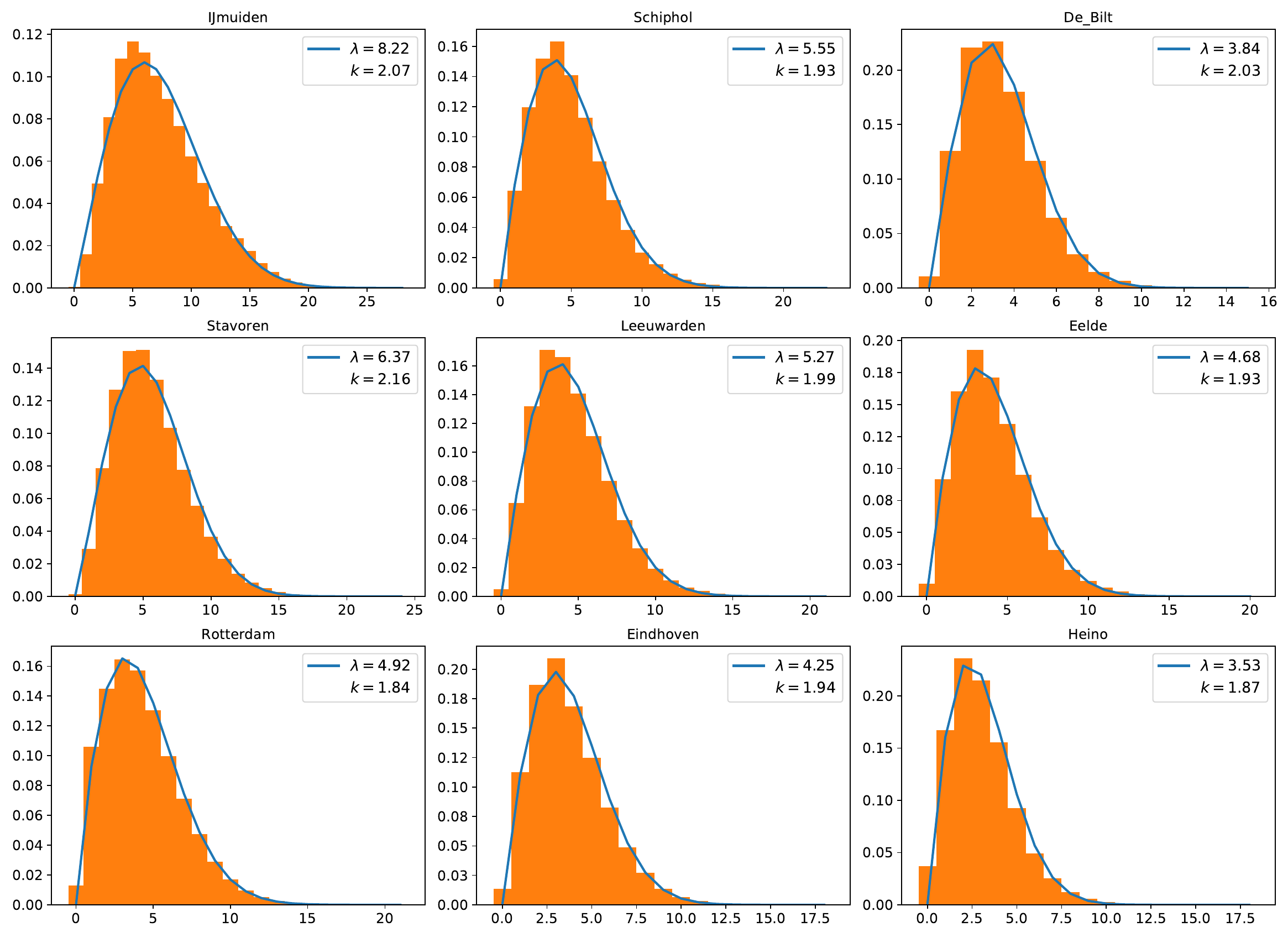}

	\vspace*{-0.1cm}
	\caption{Empirical probability density functions of wind speed at the 9 reference sites (orange histograms). Wind velocities (x-axis) are in $m/s$. The solid lines correspond to their fit by a Weibull distribution. In the inset are reported the Weibull scale ($\lambda$) and shape ($k$) parameters.}
	\label{fig_wd}
\end{figure}

\begin{figure}[th]
	\begin{center}
	\hspace*{-0.4cm}
		\includegraphics[width=10cm,angle=0]{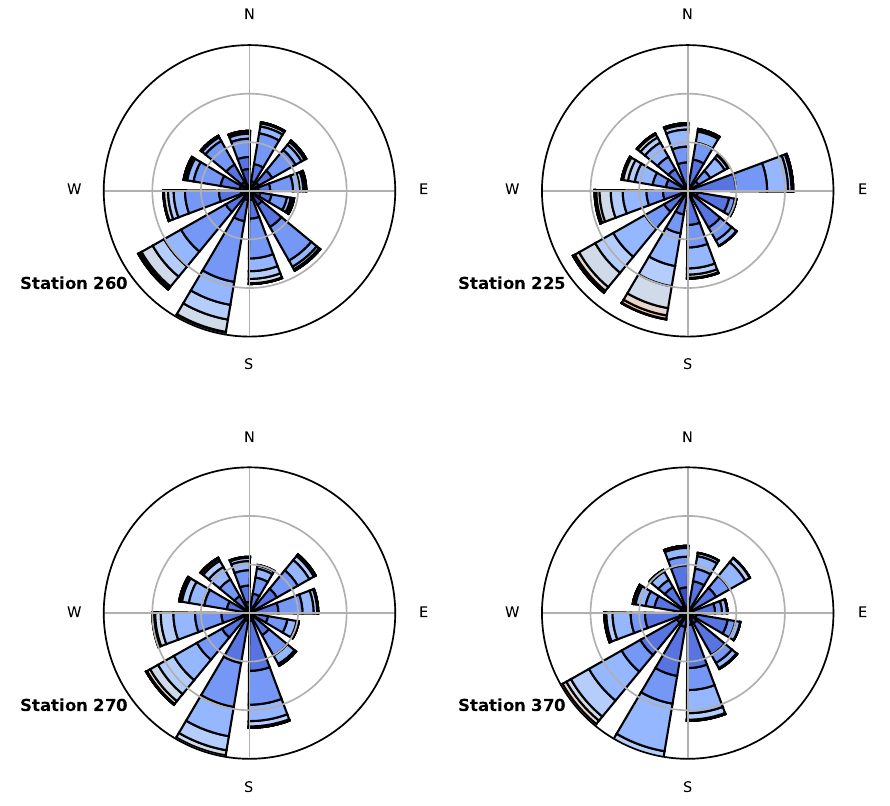}
	\end{center}
   \vspace*{-0.1cm}
	\caption{Wind roses at 4 stations from KNMI database. One can see that in each case, the main winds come from S-W direction.}
	\label{fig_windrose}
\end{figure}

The dataset we use in this paper, consists of weather data recorded by the Royal Netherlands Meteorological Institute (KNMI) at many different stations spread over the Netherlands\footnote{KNMI records are freely available online at https://www.knmi.nl/nederland-nu/klimatologie/uurgegevens.}. 
The observational network of the KNMI comprises various meteorological stations that include notably automatic stations and wind measuring masts of generally $10$ m height (see \cite{KNMI_infos} for details and also Table \ref{tab:dataset}). 
The observations have been collected during very long time periods spanning more than 70 years for some stations.
In this study, we mainly use wind data, more precisely, hourly mean amplitudes (in 0.1 $\mbox{ms}^{-1}$) and directions of the mean during the 10-minute period preceding the time of observation (in degrees) of horizontal wind.  The locations of all the 42 sites we consider as possible ``neighboring" stations of a given site are reported in Fig. \ref{fig_carte} where ($\bullet$) symbols indicate the 9 ``reference'' stations where wind speed prediction is performed. The time period we consider extends from 01-01-2001 to 12-31-2021. For all the models described below, data over the period ranging from 2001 to 2017 is used to build the training set for optimizing the model parameters while the last 3 years time intervals (2018-2021) are used as the validation set where the model performances are evaluated. The 9 reference sites of Fig. \ref{fig_carte} were chosen primarily because they have no missing data points during these periods. In table \ref{tab:dataset}, we report, respectively, the mean speed value, the maximum (hourly mean) speed values and the number of hourly samples in the training and in the testing period. We see that coastal sites are characterized by a larger mean and extreme wind speed (which can reach almost $30$ m/s) than those further west. As far as the shape of the wind speed distributions is concerned, as illustrated in Fig. \ref{fig_wd}, one can see that they can be described by a Weibull distribution with a shape parameter close to $k = 2$ and, in agreement with the previous remark, a scale parameter larger for coastal stations.
Concerning the wind direction,  in Fig. \ref{fig_windrose}, we see that the prevailing wind regime comes from the South-West.

In addition to wind speed data, other weather variables (notably temperature, pressure, and dew-point temperature) are also available at the 9 reference stations. This is generally not the case for the remaining 33 stations represented by the symbol ($\blacktriangledown$) in Fig. \ref{fig_carte} and which can be used as model input. For these sites, one can often have some periods of missing wind data. Let us specify that when there is a lack of data on a given site used as input to the model during a given hourly period, whether in the training or validation sample, we delete all the data corresponding to this slice of time. Doing so, in the worst case, i.e., when one considers {\em simultaneously} all wind data from all 42 stations, the total number of missing data is less than 15\% of the full sample. It means that the number of hourly periods with no missing data in all stations, represents more than 85\% of the total possible number of data (more precisely, in that case, one has $N_s = 136100$ samples for the training period and $N_s = 21600$ for the validation period). 
 
\section{Regression Methods}
\label{sec:reg}
As said in the introductory section, we want to compare 
various methods optimized to forecast wind speed from 1 to a few hours ahead.
All of the methods we consider belong to the general class of ``machine learning" methods since they consist in directly implementing a way of performing the best prediction from a given set of explanatory variables also referred to as ``input features" or ``input data''. It is noteworthy that we don't try to build any dynamical model, as, e.g. an ARIMA time series model, which would be calibrated on data and which would aim for some explanatory abilities. Instead, we directly tackle the problem of the best prediction, i.e., a regression from these data. Such a difference in their perspectives mainly distinguishes, within statistical approaches, statistical modeling from machine learning.

\subsection{Some definitions and notations}
Before describing the different methods we experiment with, let us introduce some notations. 
In the following $V^i_t$ ($V^i_t > 0$) and $\varphi^i_t$ ($0 \leq \varphi^i_t < 2 \pi$) will stand for respectively 
the hourly mean amplitude and the last 10-min mean direction of the velocity at time $t$ and for station $i$. The column vectors associated with all $N$ stations considered in our study ($N=42$ as indicated in Fig. \ref{fig_carte}) will be denoted as $\bV_t$ and $\bF_t$. We denote by $u^i_t$ and
$v^i_t$ the Cartesian components of the surface wind velocity at station $i$, i.e., $u^i_t = V^i_t \cos(\varphi^i_t)$ and $v^i_t = V^i_t \sin(\varphi^i_t)$ and then by $\bu_t$ and $\bv_t$ the corresponding N-dimensional vectors associated with all stations. In various prediction methods, it may also be interesting to account for seasonal (i.e. annual) and also for diurnal (i.e. daily) variations in wind regimes. For that purpose, if $H_t \in [1 \ldots 24]$ and $D_t \in [1 \ldots 365]$ denote respectively the hour of the day and the day of the year associated with time $t$, we define the vector of parameters $\bS_t$:
\be
\label{eq:defS}
\bS_t = \left.
\begin{bmatrix}
	\cos(\frac{2 \pi H_t}{24}) \\
	\sin(\frac{2 \pi H_t}{24})  \\
	\cos(\frac{2 \pi D_t}{365}) \\
	\sin(\frac{2 \pi D_t}{365})  
\end{bmatrix}
\right. \; .
\en

If $\bX^{(1)}_t$, $\bX^{(2)}_t$,$\ldots$, $\bX^{(M)}_t$ are $M$ vectors of arbitrary dimensions $(N_1,N_2, \ldots,N_M)$ at time $t$, 
we define the following vector of dimension $\sum_k N_k$ :
$$
(\bX^{(1)},\bX^{(2)},\ldots,\bX^{(M)})_{t} = \left.
\begin{bmatrix}
	\bX^{(1)}_t \\
	\bX^{(2)}_t \\
	\vdots \\
	\bX^{(M)}_{t} 
\end{bmatrix}
\right. \; .
$$
On the same ground, for any time dependent vector $\bX_t$, 
we will denote by $(\bX)_{t,n}$ the $n N$-dimensional vector of lagged vectors:
$$
(\bX)_{t,n} = \left.
\begin{bmatrix}
	\bX_t \\
	\bX_{t-1} \\
	\vdots \\
	\bX_{t-n+1} 
\end{bmatrix}
\right. \; .
$$
Accordingly, we can define the vector of lagged components:
$$
(\bX^{(1)},\bX^{(2)},\ldots,\bX^{(M)})_{t,n} = \left.
\begin{bmatrix}
	(\bX^{(1)})_{t,n} \\
	(\bX^{(2)})_{t,n} \\
	\vdots \\
	(\bX^{(M)})_{t,n} 
\end{bmatrix}
\right. \; .
$$

All the methods defined below can be loosely considered as aiming to solve the following statistical regression problem:
Let $\bX_{t}$ be a $J$-dimensional vector of explanatory variables (for example, $\bX_{t} = (\bV, \bF)_{t,n}$ built from 
the $J = 2 N n$ lagged wind amplitudes and directions as defined previously) and consider $F_\bT(\bX)$
a given class of functions $\RR^{J} \rightarrow \RR^{O}$ characterized by the set of parameters $\bT$. Let $\bY_t$ be some $O-$ dimensional 
observable vector  (for example, if $O = 2$, $\bY_t$ can be the 2-dimensional vector of wind speed values at 2 given stations $i_1$ and $i_2$.) ;
one wants to find $\bT_0$ such that the observation $\bY_{t+h}$ at an horizon $h$ ahead in the future, can be written as
\begin{equation}
\label{nlr}
  \bY_{t+h} = F_{\bT_0} \big( \bX_{t} \big) + \bm{\varepsilon}_{t+h}
\end{equation}
where $\bm{\varepsilon}_{t+h}$ is a $O$-dimensional ``noise'' error vector. In practice, one seeks for $\bT_0$ such that:
\begin{equation}
\label{eq:loss}
 {\bT}_0 = \argmin_{\bT}  \E \Big[ {\mathcal E} \Big( \bY_{t+h}, F_{\bT} \big( \bX_t \big) \Big) \Big]  + P \big( {\bT} \big) \; .
\end{equation}
where $\E$ stands for the expectation as respect to the law of $\{\bY_t\}_t$ and $\{\bX_t\}_t$, ${\mathcal E}(x,y)$ stands for an ``error''
between $x$ and $y$.
The so-defined best prediction of $\bY_{t+h}$ is usually denoted as ${\widehat \bY}_{t+h}$, i.e., 
\begin{equation}
\label{eq:haty}
	  {\widehat \bY}_{t+h} = F_{\bT_0} \big( \bX_{t} \big) \; .
\end{equation}
The error function in Eq. \eqref{eq:loss} is commonly chosen to be $\ell_1$-norm (i.e. the sum absolute difference of coordinates) 
${\mathcal E}(x,y) = \lVert x-y\rVert_1$ or the $\ell_2$-norm (i.e., the square of the Euclidian distance) ${\mathcal E}(x,y) = \lVert(x-y)\rVert_2^2$. In that case $\E \left[ {\mathcal E}(x,y) \right]$ is referred to as the ``Mean Square Error" (MSE). $P(\bT)$ in Eq.  \eqref{eq:loss} stands for a regularization term in order to eventually constrain the solution (e.g., $P \big( {\bT} \big) = \lambda \lVert\bT\rVert_2$). For the purpose of this study, we choose to focus exclusively on the mean-square error and we disregard any penalization term. It is noteworthy that, from a theoretical point of view, the best function $F(\bX_t)$ that minimizes the MSE is the conditional expectation $\E(\bY_{t+h} | \bX_t)$
and in that respect, any model $F_{\bT} \big( \bX_t \big)$ that targets the smallest MSE, tries to approximate this conditional expectation as well as possible .

All the prediction methods we consider in this paper enter in this framework with 
\be 
\label{eq:def_Y}
\bY_{t+h} = \left.
\begin{bmatrix}
	V^{i_1}_{t+h_1} \\
	V^{i_2}_{t+h_2} \\
	\vdots \\
	V^{i_O}_{t+h_O} 
\end{bmatrix}
\right. \; .
\en
where, for any $k \in [1,\ldots,O]$, $V^{i_k}_{t+h_k}$ is the surface wind velocity at site $i_k$ and time $t+h_k$. The horizons $h_k$ are chosen to vary between 1 and 6 hours. The vector of explanatory variables (features) $\bX_t$ is built from wind velocity components of nearby stations that are scaled by an usual standardization procedure.  We will also eventually consider additional factors like the atmospheric variables (e.g., pressure, temperature) or some deterministic variables like the vector $\bS_t$ defined in \eqref{eq:defS} in order to account for possible diurnal and seasonal effects.

\subsection{Prediction methods}
In this section we describe the various prediction methods that we use or that we refer to 
in the paper.




\paragraph{Persistence}
As explained in \cite{anemos1}, simple techniques are often
used as references within the wind speed or wind power forecasting community.
The persistence model is probably the most commonly used reference predictor and
according to Giebel \cite{giebel},
for short prediction horizons (from a few minutes to a few hours),
this model is the benchmark all other prediction models have to beat.
The persistence model simply consists in assuming that the best prediction is provided by the latest observation, i.e., one chooses 
$\bX_t = \bY_t$ and
\begin{equation}
\label{eq:pers}
  F^{\mbox{\small{pers}}}(\bf Y_t) = \bY_t \; .
\end{equation}
From a mathematical point of view, it amounts to assuming that the conditional expectation $\E(\bY_{t+h} | \bY_t) = \bY_t$ and therefore that $\bY_t$ is a martingale. In section \ref{sec:results}, this method will be used as a simple benchmarck in order to compare the improvements of different approaches.

\paragraph{Linear regression}
The second class of models we consider is also among the simplest ones, namely the class of  linear predictions where the function $F$ in Eq. \eqref{nlr} is the following 
affine function: 
\begin{equation}
\label{eq:mod_linear}
   F^{\mbox{\small{Lin}}}(\bX_t)  = \bA \bX_t + \bB  \;, 
\end{equation}
where $\bX_t$ is the $J$-dimensional vector of explanatory variables and $\bT = (\bA,\bB)$ are respectively a fixed $O \! \times \! J$ matrix and a fixed $O$-dimensional vector. As recalled in the introduction, linear prediction models have been considered in many former studies and notably few of them include spatially-distributed wind speed as explanatory variable $\bX_t$ \cite{Lars2006,Hill2011,Filik16}.
In this work, $F^{\mbox{\small{Lin}}}(\bX_t)$ as defined in Eq. \eqref{eq:mod_linear} is directly implemented using the ``LinearRegression'' model from the Python Scikit-Learn library \cite{Pedregosa:2011}. 
\begin{figure}[!th]
	\centering
	\vspace*{-3cm}	
	\hspace*{-0.5cm}
	\includegraphics[width=10cm,angle=0]{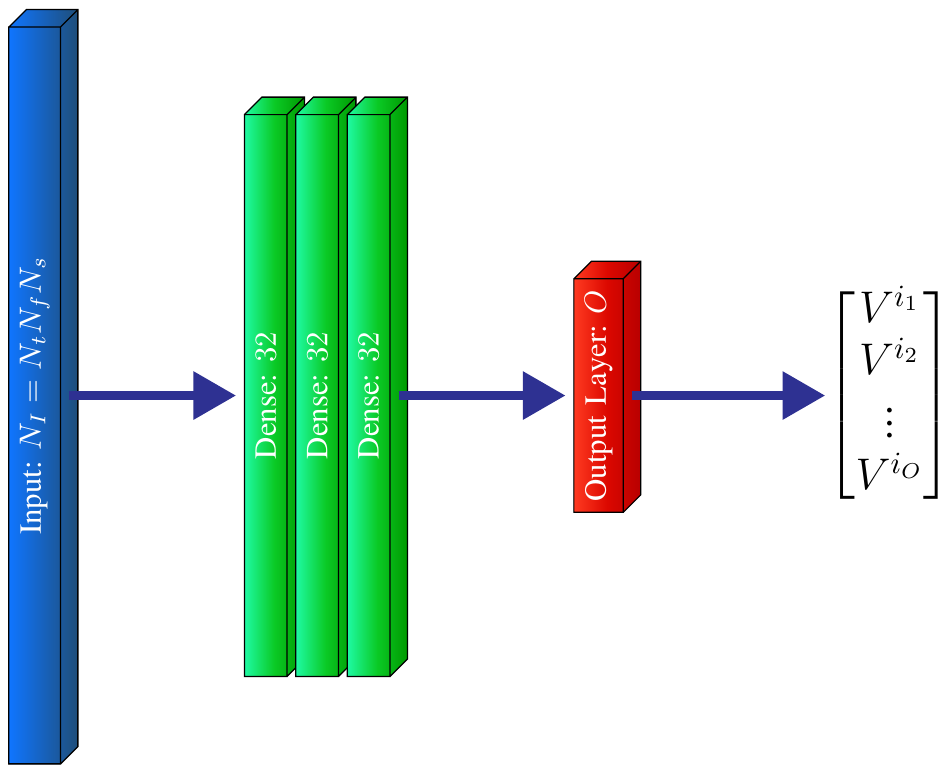}
	
	\vspace*{-3cm}
	\caption{The simplest DNN architecture is made of 3 fully connected (``dense'')  layers, each with 32 output channels and a ReLU activation function.}
	\label{fig_dense}
\end{figure}

\begin{figure}[!th]
	\centering
	\vspace*{-3cm}
	\hspace*{-0.5cm}
	\includegraphics[width=10cm,angle=0]{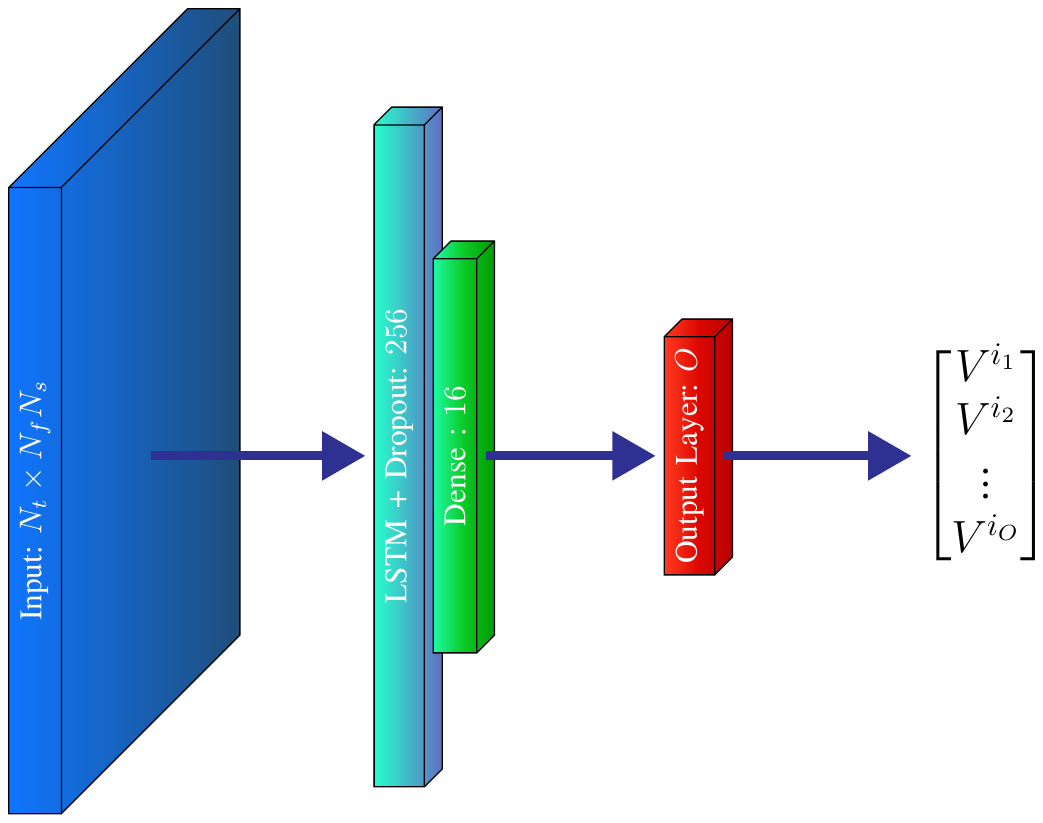}
	
	\vspace*{-3cm}
	\caption{The LSTM model we considered is made of one LSTM layer with 256 output channels followed by a $16$ unit dense layer.}
	\label{fig_lstm}
\end{figure}

\begin{figure}[!th]
	\centering
	\vspace*{-4cm}
	\hspace*{-0.7cm}
	\includegraphics[width=10cm,angle=0]{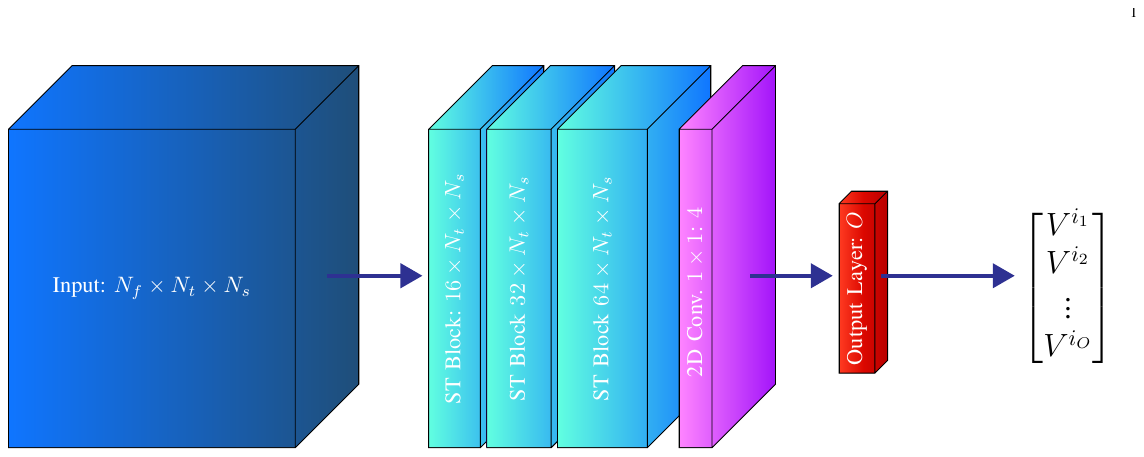}
	
	\vspace*{-5cm}
	\caption{This figure represents a sketch of the Graph Convolutional Neural Network proposed in Ref. \cite{gcn_ws_2021}.}
	\label{fig_gcn}
\end{figure}

\begin{figure}[!th]
	\centering
	\vspace*{-3cm}
	\hspace*{-0.5cm}
	\includegraphics[width=10cm,angle=0]{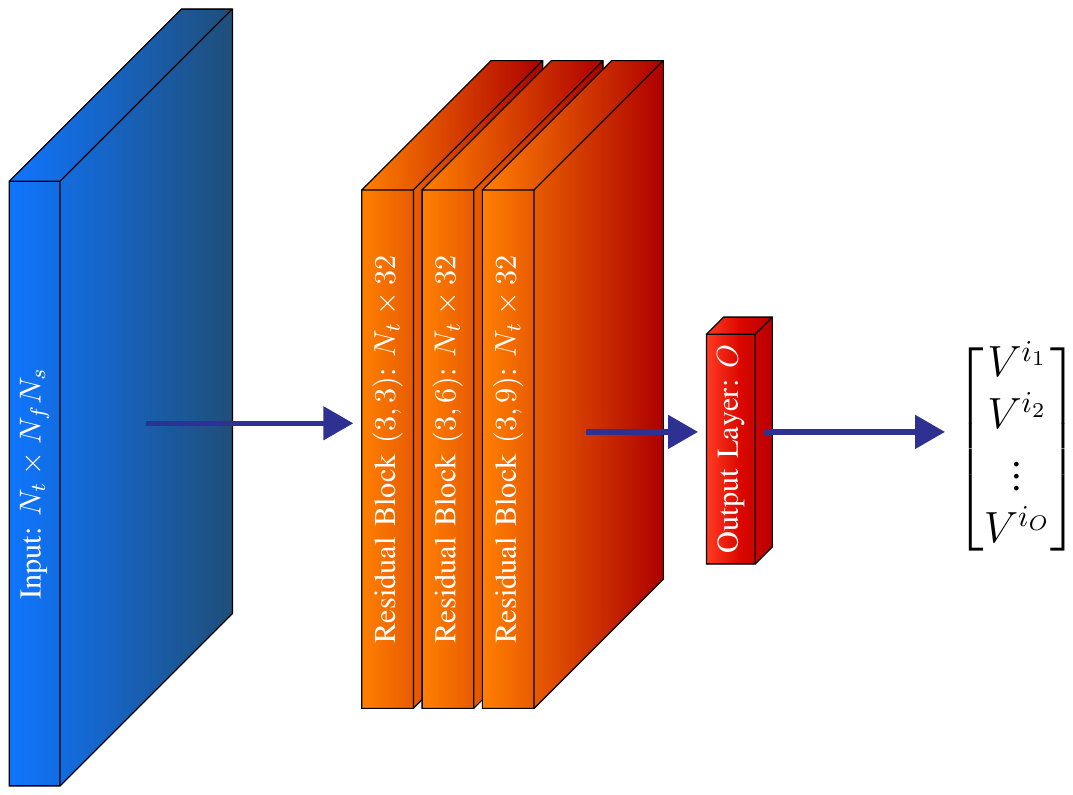}
	
	\vspace*{-3cm}
	\caption{Temporal Convolution Network made of 3 ``residual blocks'' each with 32 output channels, a convolution kernel of size $3$ and dilation factors that are respectively $3$,$6$ and $9$.}
	\label{fig_tcn}
\end{figure}
\paragraph{Gradient boosting regression trees}
Boosting methods consist in aggregating various (``weak'') learners in an iterative way such that every additional method  is optimized to correct the deficiencies of the previous ones in order to achieve an overall
better regression \cite{hastie01}. 
A gradient boosting method is a boosting method where the optimization is performed as a gradient descent in a space 
of functions (instead of parameter space) that corresponds to the space of weak learners. When the latter are regression trees, the method is called ``Boosted Regression Trees'' or, in short, Gradient Boosting (GB). 
Therefore, a GB method mainly consists in getting the best order $R$ regression function
$F_R (\bX_t)$ which corresponds to the summation of $R$ regression-trees: 
$$
F_R^{\mbox{{\small GB}}}(\bX_t)  = \sum_{i=1}^R f_i(\bX_t)
$$
where every $f_i(\bX_t)$ is a regression-tree of fixed depth $D$, namely a piece-wise constant function defined over a partition of the $J$ dimensional hypercube representing the domain of $\bX_t$. 
Gradient boosting method represents a very flexible non-parametric machine learning technique for regression (or classification) that has proven to be very efficient in many contexts. Although its use in wind speed or wind power forecasting has been limited (see e.g., \cite{xgb_wind2020,xgb_wind2021}), given its very competitive performances in many other domains, it deserves to be considered in the present study. XGBoost \cite{xgboost_2016} and LightGBM \cite{lgbm_2017} are two of the most popular GB algorithms that achieve state-of-the-art performances for fitting the relationship between features and labels.
Even though we implemented and tested both methods, since we obtained results that are very close, we only reported the ones obtained with LightGBM.

\paragraph{Deep neural networks}

Among all machine learning methods used to address wind speed forecasting problems, as reviewed in the introduction, 
Deep Neural Networks (DNN) are undoubtedly the most widely used and the most efficient methods.  
Let us recall that an artificial neural network roughly consists 
in describing the non-linear relationship $F_{\bT} \big( \bX_{t} \big)$ in Eq. \eqref{nlr} between the input $\bX_t$ and the output $\bY_{t+h}$, as the composition of $L \geq 1$ intermediate non-linear functions (the first one is referred to as the ``input layer", the other ones are called ``hidden layers'') and a final output function (the ``output layer''). Each layer is made of a fixed number of units, the ``neurons'' that perform a linear transformation and then apply a non-linear
``activation'' function common to all units of a given layer. One generally speaks of ``Deep'' neural network when the number of hidden layers is large enough, say $L \geq 2$ (see e.g., \cite{Good16}). There are many different types of neural networks and in this work, our goal is resolutely not to propose any new architecture that would perform better than existing ones or even to finely tune any existing method. We rather consider different DNN models, more or less simple, that are either generic for time series forecasting tasks or have been shown to be effective in the case of wind speed forecasting. Our objective is to compare them with each other or with linear and boosting methods described previously. 
More specifically, we will consider four different DNN models: First, we will test the class of ``fully connected'' (or ``dense'') architectures where, for each layer, every unit takes as input the output of all the previous layer units. 
The precise model we consider is depicted in Fig. \ref{fig_dense} where 3 dense layers, each with 32 output units and a ReLU activation function, are stacked before the output layer. The second class of models we consider is a so-called recurrent neural network that has proven to achieve close to state-of-the-art performances in many time series forecasting problems. We use a very simple model involving a so-called Long Short Time Memory (LSTM) recurrent layer (the most widely used variant of RNN) with 256 output channels that feed a dense layer of 16 output channels before the final layer (see Fig. \ref{fig_lstm}). Finally, the two last networks that we consider, are mainly convolutional neural network models. We consider the 
so-called ``WeatherGCNet'' which is a graph convolutional neural network introduced in \cite{gcn_ws_2021} which has proven to be very efficient for wind speed prediction precisely on the same Netherlands wind speed dataset described in section \ref{sec:data}. As illustrated in Fig. \ref{fig_gcn}, this network mainly involves 3 ``Spatio-Temporal" (ST) blocks that are themselves made of a graph convolution layer with a learnable adjacency matrix followed by a temporal convolution layer.
We refer the reader to Ref. \cite{gcn_ws_2021} for further details. Finally, we also experiment with a model based on Temporal Neural Network as introduced in Ref. \cite{tcn1}. This model, illustrated in Fig. \ref{fig_tcn}, mainly consists in stacking  
``residual blocks'' of various dilation factors, each block is essentially made of 2 causal temporal dilated convolution layers and a skip connection (see \cite{tcn1} for a detailed description). As emphasized in \cite{tcn1} for many different tasks and in \cite{tcn2} specifically for time series forecasting, TCN performances can exceed that of RNN models.

\subsection{Performance analysis metrics}
In order to assess the performances of different wind speed prediction methods, we refer to the most commonly used metric, namely the Mean-Square-Error (MSE):
\be
\mbox{MSE} = \frac{1}{n}\sum_{t=1}^{n_T} \sum_{i=1}^O \left( Y^i_{t+h} - {\widehat Y}^i_{t+h} \right)^{2}
\en
or its square root, the Root-Mean-Square-Error (RMSE),
\be
\mbox{RMSE} = \sqrt{\mbox{MSE}} \; ,
\en
where $O$ is the dimension of the output vector and $n_T$ stands for the number of observations in the considered period (training or validation).
MSE precisely corresponds to the cost function ${\mathcal E}(x,y)$ we choose for the model parameter optimization as defined in Eq. \eqref{eq:loss}. 
\begin{figure}[!ht]
	\begin{center}
		
		\resizebox{\columnwidth}{!}{%
		\includegraphics[width=17cm,angle=0]{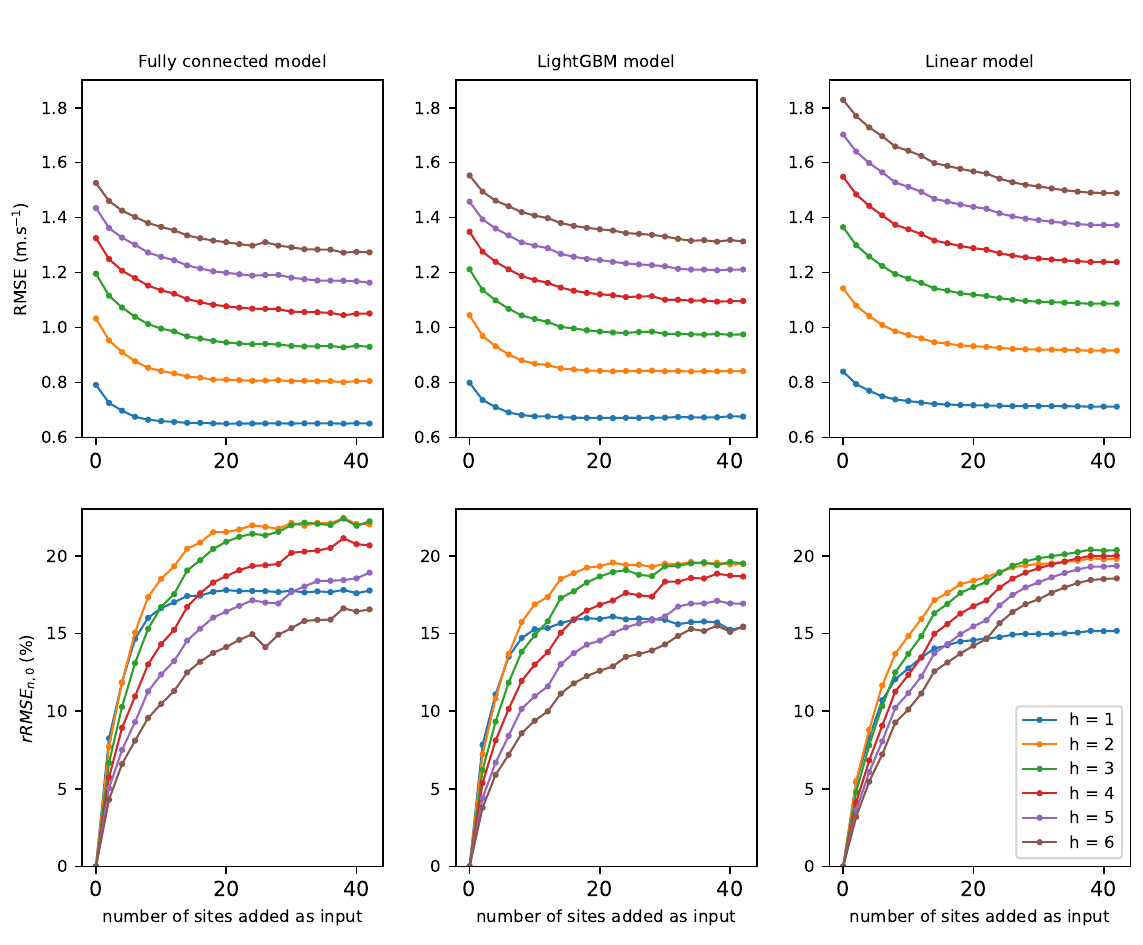}
         }		
	\end{center}
    \vspace*{-0.5cm}
	\caption{Evolution of RMSE values obtained with 3 different methods : a fully connected model at left, a lightGBM at the middle and a linear model at right. Results for different forecast horizons are represented, from 1 to 6 hours, as a function of the number $n_N$ of the closest neighboring sites added as input of the model. In the figures at the top, we represent the mean of the RMSE obtained with the 9 sites presented above, for each forecast horizon. At the bottom, we have represented $rRMSE$ that allows one to compare the RMSE obtained by adding sites as input to the model (version of the model one wants to test), with respect to the RMSE obtained by the same model, with only the studied site as input (reference version). For each forecast horizon, we can see that RMSE decreases when the number of input neighbors increases, until reaching a plateau all the more quickly as the horizon is short.}
	\label{fig_addsites}
\end{figure}

In order to compare the performance of a test model ($mod$) against some {\em reference} model ($ref$), we use the following 
relative error:
\begin{equation}
\label{eq:def_rRMSE}
\mbox{rRMSE}_{mod,ref} = 100  \left( 1- \frac{\mbox{RMSE}_{mod}}{\mbox{RMSE}_{ref}} \right)
\end{equation}
which gives the percentage of improvement of the {\em test} forecasting method as compared to the reference (RMSE$_{ref}$ is the RMSE obtained with the reference model while RMSE$_{mod}$ is the RMSE obtained with the model one wants to test).

\section{Empirical results}
\label{sec:results}
In this Section, we report and discuss the wind speed forecasting performances of the methods we presented previously. As we have already specified in the introduction, our objective in this work is not to seek the very best effective model, specifically optimized for predicting the wind velocity amplitude at some given location and horizon ahead but rather to uncover some ``stylized facts'', to discuss on a general ground which features and model properties are important and to what extent. Therefore, in order to avoid a cumbersome fine-tuning procedure, most of the model hyperparameters were either chosen to correspond to the usual practices found in the literature or chosen within a small number of possibilities around these values. We observed that changes within the range of values commonly considered by the community, only slightly affect the overall performances reported below. The vectors of input features chosen for each of the prediction models and their main hyperparameters are reported in Table \ref{tab:hyper}. As mentioned in Sec. \ref{sec:data}, the model parameters are optimized over a training period extending from 01-01-2001 to 12-31-2018 and the model performances are estimated over a validation period from 01-01-2019 to 12-31-2021. The final MSE we report corresponds to the average value of the MSE obtained for each of the 9 reference stations shown in Fig. \ref{fig_carte}. 
   
\begin{table*}[t]
	\centering
   \resizebox{\columnwidth}{!}{%
	\begin{tabular}{|l||l|l|l|l|l|l|}
		\hline
		\multicolumn{1}{|c||}{Parameters / Models}         & \multicolumn{1}{|c|}{Linear} &
		\multicolumn{1}{|c|}{Dense} & \multicolumn{1}{|c|}{LSTM}        & \multicolumn{1}{|c|}{GCN}         &
		\multicolumn{1}{|c|}{TCN}  & \multicolumn{1}{|c|}{LGBM}  \\ \hline \hline
		Input Features & $(\bV)_{t,n}$ & $(\bu,\bv,\bS)_{t,n}$ & $(\bu,\bv,\bS)_{t,n}$ & $(\bu,\bv,\bS)_{t,n}$ & $(\bu,\bv,\bS)_{t,n}$ & $(\bu,\bv,\bS)_{t,n}$ \\ \hline
		Order $n$ of used past data &  $3 \times h$ &  $3 \times h$ &   $3 \times h$   &   $3 \times h$    &  $3 \times h$ &  $3 \times h$    \\ \hline 
		Num. of hidden layers & - & 3 (Dense) & 2 (LSTM + Dense) & 3 (ST blocks) & 3 (Res. blocks) & -  \\ \hline 
	    Num. of layer channels &  - & (32,32,32) & (256,16) & (16,32,64) & (32,32,32) & -      \\ \hline
		
	    Num. of leaves &- & - &  - &  - &  - & 16    \\ \hline 
		Max. depth &- & -  &  -    &  -     &  - & 5    \\ \hline 
		Num. estimators &- & - &  -     &  -   &  - & 1000     \\ \hline 
		
	\end{tabular}
   }  
	\vspace*{0.5cm}
	\caption{Input features and main hyperparameters for each forecasting model. The learner used for each DNN was ADAM with default parameters and with a mini-batch size of $512$ associated with an early stopping with patience of $20$ steps. Each velocity component of the input features is centered by its mean value and scaled by its standard deviation.}
	\label{tab:hyper}
\end{table*}

\subsection{Improvement obtained by using surrounding site wind data.}
First of all, we would like to quantify the benefit of using, as input to a given model, all wind speed data from nearby stations. For that purpose, for all of the 9 reference sites of our database, we estimate the improvement of the forecasting performances when one takes into account, incrementally, the wind speed components of the closest stations: we first add, in the model input data, the velocity components of the nearest site, then the components of the two closest sites and so on. In Fig. \ref{fig_addsites} are reported the results we obtained for 3 different models, namely the fully connected Neural Network described in Fig. \ref{fig_dense} (left panels of Fig. \ref{fig_addsites}), a lightGBM gradient boosting model (middle panels of Fig. \ref{fig_addsites}) and a linear model (right panels of Fig. \ref{fig_addsites}). The considered forecasting horizon varies from  1 hour (in blue) to 6 hours (in brown). The mean RMSE of each model and each horizon is reported in the top panels as a function of the number of considered closest neighboring stations which varies between $n_N = 0$ (no neighboring site is considered) and $n_N = 41$ (all available sites are considered). One clearly sees that all curves are decreasing meaning that, whatever the considered model and whatever the forecasting horizon, 
using more and more wind speed data from neighboring locations significantly reduces the forecast RMSE error. We can also remark that among the 3 tested models, the Neural network model provides the best forecasting performance while the linear model is, by far, the worst one. In order to better quantify the relative gain obtained by considering "off-site" wind data, in the bottom panels of Fig. \ref{fig_addsites}, we have represented $rRMSE_{n,0}$ as defined in Eq. \eqref{eq:def_rRMSE} where the reference model is the model prediction when we don't account for the neighboring stations ($n_N=0$) and the test model is the one that uses the $n_N = n$ closest station data. In that respect, this quantity directly gives the percentage of the performance improvement when one increases $n$, the number of neighboring site data as input (as defined previously for the top figures). We can see that for all models and all horizons, the gain increases: It approaches or even exceeds 20 \% in the case of the fully connected neural network. It is clear that accounting for the wind velocity of neighboring stations allows one to capture spatiotemporal information. In particular, sites in the upwind direction are expected to be the most relevant for anticipating future weather conditions since they are, in a way, transported by the main surface wind. This can explain why, when $h=1$ hour, one observes a quick saturation, around $n_N=10$, of the rRMSE (blue lines in the bottom panels): there is no point in adding as input, data from sites too far away to provide useful information on such a short horizon of time.
Since neighboring sites are added incrementally from nearest to farthest,
we can expect such a plateau to be observed for $n_N$ all the larger as the forecast horizon is large. For the 2 hour forecast horizon (orange line), it occurs when $n_N=20$ sites are added as input while for largest horizon, farthest stations are not sufficiently distant to observe any constant behavior. Indeed,  it is not $n_N$, the number of added neighboring stations, that is real important parameter but rather the distance between the added site and the studied one ; the longer the forecast horizon, the more interesting it is to consider distant sites as input to the model. Let us remark however that, when $h=1$, the maximum theoretical improvement appears to be significantly smaller than for larger horizons. This may be due to various reasons like the granularity of the considered horizons and of the spatial distribution of the sites around a given location. One can also expects a smaller potential improvement (and thus a smaller rRMSE) at horizon 1 because persistence is more likely to contribute at small time scales. Addressing such questions related to the precise shape of the error curves as a function of the horizon and the number of surrounding sites, would require further numerical experiments and significantly more data. 

\begin{figure}[th]
	\begin{center}
		\hspace*{-0.5cm}
		\includegraphics[width=10cm,angle=0]{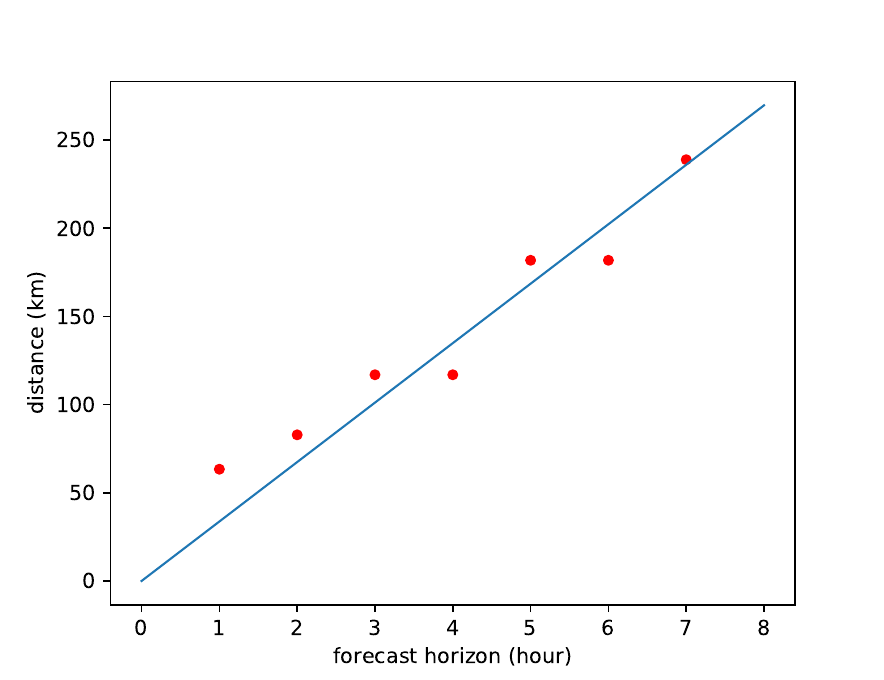}
	\end{center}
	\caption{For each forecast horizon $h$ from 1 to 7 hours, one has reported the distance $D$ between the studied station ``Leeuwarden" (270) and the site added as input and which gives the smallest RMSE when the wind comes from the South-West. One roughly observes a linear relationship $D = V_0 h$ with $V_0 \simeq 35$ km/h. }
	\label{fig_dist}
\end{figure}

In order to empirically confirm the previous assertion and validate the idea that wind is advected by itself so that spatial information can significantly improve wind speed prediction, we choose to study, for a given site, how far it is pertinent to fetch data for a given forecast horizon. For that purpose, we focus on the wind speed forecast at the station ``Leeuwarden" (270) located in the North of the Netherlands. As it can be checked on the map of Fig. \ref{fig_carte}, this station has its neighbors spread over a wide range of distances towards the South-West that corresponds precisely to the direction of the prevailing wind regime (see Fig. \ref{fig_windrose}). One after the other, the wind data from each site at S-W of ``Leeuwarden" are considered, together with the local wind, as the input to the forecasting model. Let us mention that in order to increase the sensibility of our experiment, we chose to estimate the forecast RMSE restricted to the periods when the wind direction is from S-W, namely within a $30^o$ S-W sector. In short, we consider the model forecasting performance when the wind is S-W and when one adds pieces of information from a single distant station along this upwind direction. In Fig. \ref{fig_dist}, we have represented, for each forecast horizon from 1 to 7 hours, the distance between ``Leeuwarden" station and the added station that provides the best prediction when using the Fully Connected Neural Network (other methods lead to the same kind of results). We see that the longer the forecast horizon, the more relevant it is to seek information from remote sites. One can also see that, even if the values of distances are poorly sampled, a linear regression can be performed and corresponds to advection of information by a mean wind speed of around $35$ km/h. Let us notice that this value is very large as compared to the mean wind speed in that direction, which is close to $15$ km/h. This might be explained by the fact that the model's largest errors are obtained when one observes strong wind conditions. A precise answer to this question will be considered in future work.

\begin{figure}[!ht]
	\begin{center}
		\includegraphics[width=15cm,angle=0]{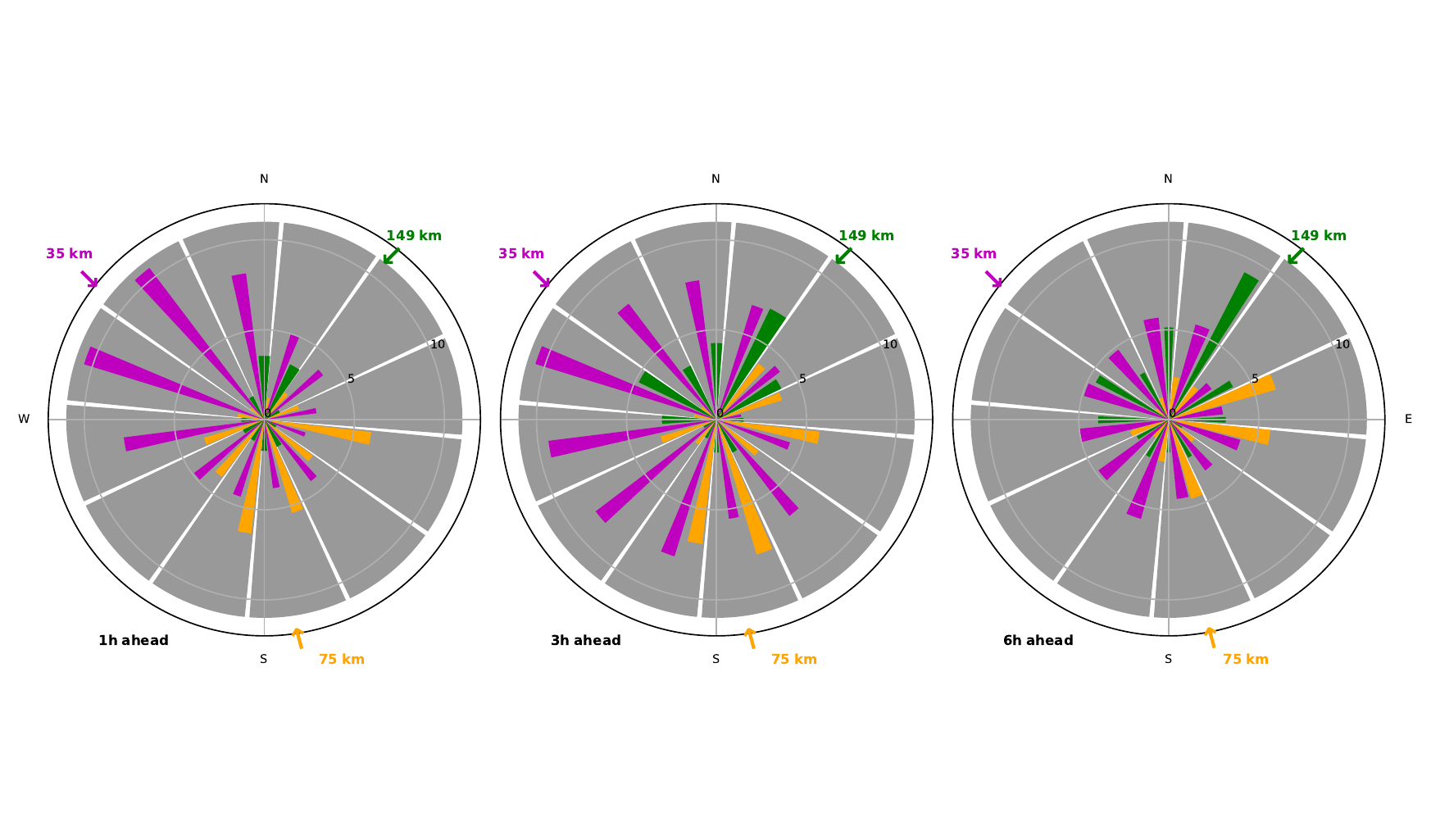}
	\end{center}
\vspace*{-2cm}
	\caption{rRMSE between two configurations of the fully connected model for the station 260 ``De Bilt'', depending on the wind direction measured at this site : first, in the reference model, we forecast the wind speed at the site 260 with only data from site itself as input ; second, for the test model, we forecast the wind speed at the site 260 by also adding as input wind data from one neighboring station. In each figure (1 hour horizon to the left, 3 hours in the middle and 6 hours to the right), the colored arrows give the direction and the distance of the site added as input. The length of the different color bars indicates the relative improvement (from zero to 10 \%) as compared to the reference model version.}
	\label{fig_dir}
\end{figure}


As well as the distance, we can also study how the direction of the site one adds as input to the model, can influence the forecasting performance. 
To quantify the impact of using the wind speed of 
a single station (say $i$) as an additional feature in order to predict the wind at some other station (say ``De Bilt") as a function of its direction, one can consider the relative score rRMSE$_{mod,ref}$ where the reference model is the ``raw-model" and the test model is the model with additional data from the site $i$.  
This score directly quantifies, in percentage, how much the added site in input improves the forecast, depending on the wind direction measured at the studied station. In the 3 panels of Fig. \ref{fig_dir}, we have displayed such rRMSE$_{mod,ref}$ obtained using the fully connected model, at the station $260$, for 3 different horizons, from 1 hour (at left) to 6 hours (at right). In these figures, the rRMSE is estimated as a function of the direction of the wind measured at the studied site, the wind direction being distributed in angles of opening $30^o$. In each panel of Fig. \ref{fig_dir}, the colored arrows indicate the direction of the used neighboring site $i$ with respect to station De Bilt.
We can observe that adding a site as input to a model allows one to improve the forecast, especially when the wind direction measured at the studied station is close to that of the added site as input, in other words, when the wind comes from the related region. In agreement with the previous discussion, the shorter the forecast horizon, the closer the added site must be in order to improve the forecast. Similarly, the more distant the added site, the less the effect on improving the forecast will be if the horizon is short.

\subsection{Model performances comparison and variable selection issues} 

Our previous results have shown that, to obtain accurate wind speed forecasts, it is important to account for the wind speed at distant stations located upwind. One can naturally wonder which of the models presented in 
Sec. \ref{sec:reg} has the greatest capacity to exploit this spatiotemporal information and also what is the level of the improvement one can expect as compared to the ``straightforward'' persistence model (Eq. \eqref{eq:pers}). For that purpose, we compute, at horizons 1, 3 and 6 hours, the RMSE (averaged over the 9 sites mentioned previously) for the persistence method and six machine learning models described in Sec. \ref{sec:reg}: the linear model, a fully connected DNN model, a GCN model, a LSTM model, a TCN model or a LightGBM approach. For each of these models, we considered the same input data, namely the last $3 \times h$ wind speed data measured at the considered site and at the 20 nearest neighboring sites.  We considered only the 20 closest stations because, as one can see in Fig. 7, whatever the model considered, the improvement when going from 20 to 40 neighboring sites is (at best) less than $4$ points. As previously emphasized, this is because for the shortest time horizon (1 hour) the additional information is useless (the additional sites are too far away) while for the largest time horizon (6 hours) they are probably too close to provide any significant gain. In that respect, we consider that 20 sites is enough to compare different methods and different models and going from 20 to 40 sites would only change the results very marginally.

\begin{figure}[!th]
	\begin{center}
          \hspace*{-0.5cm}
          \includegraphics[width=18cm,angle=0]{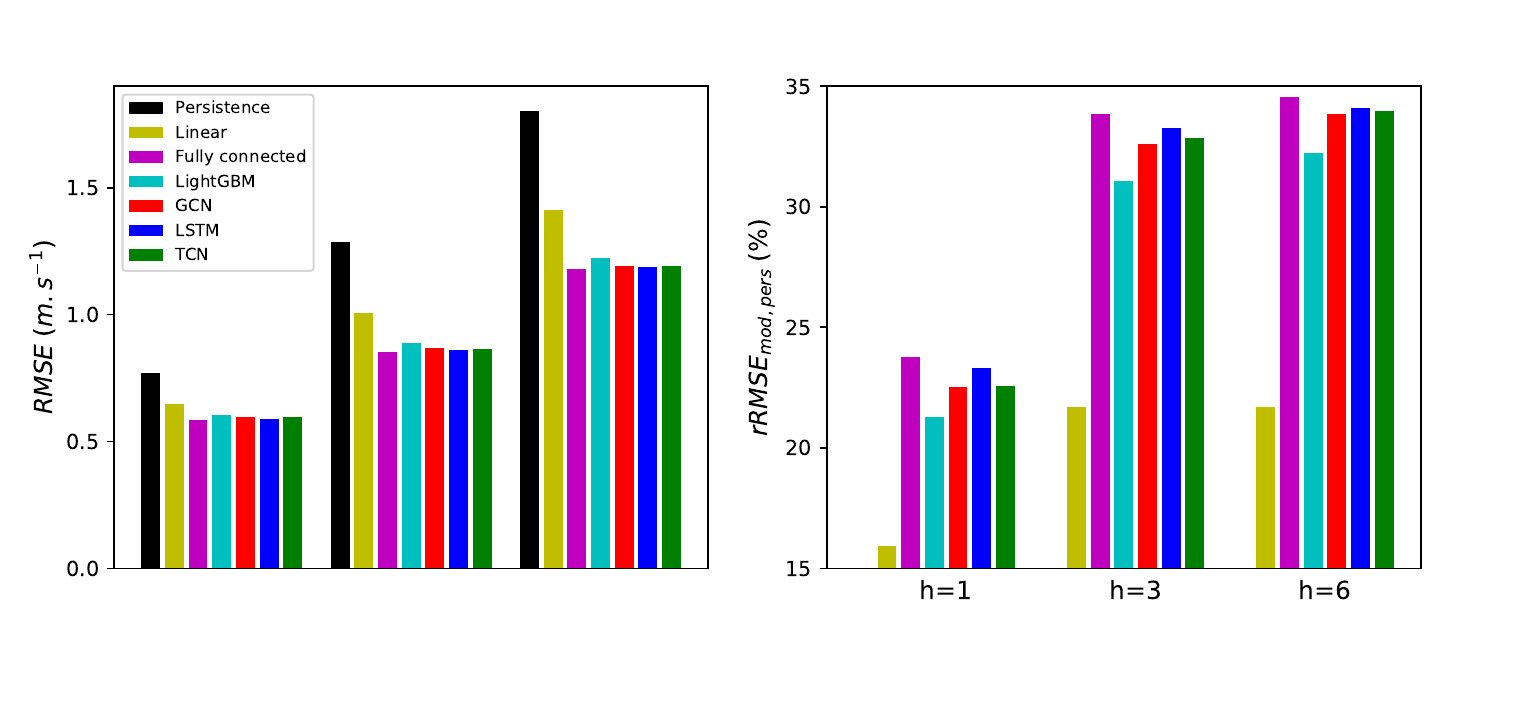}
	\end{center}
  \vspace{-2cm} 
	\caption{(Left Panel) Average of the RMSE values  (in $ms^{-1}$) obtained independently for the nine reference sites and using different prediction models, the persistence and six ML methods (namely, a linear model, a fully connected model, a LightGBM model, a GCN model, a LSTM model and a TCN approach) at three forecast horizons from 1 hour to 6 hours. Each of the 6 ML models is used with 20 neighboring sites as input.  The lower the bar, the better the forecast. (Right Panel) Average of the rRMSE values for the nine reference sites, where the reference model is persistence and the test model is one of the  six different above cited ML approaches.}
	\label{fig_compare}
\end{figure}

As one can see in the left panel in Fig. \ref{fig_compare}, all the models perform significantly better than the persistence and this is all the more true the longer the prediction horizon.
In order to better quantify the improvement of each method and to more precisely compare them to each other,
we have reported, in the right panel of Fig. \ref{fig_compare}, the rRMSE$_{mod,pers}$ score (see Eq. \eqref{eq:def_rRMSE}) with the persistence as the reference model and where the test model is one of the six alternatives listed above. We thus see that the gain of the best methods goes from more than 20 \% at horizon $h=1$ hour to more than 30 \% when $h=6$ hours. At first glance, it appears that none of the four deep neural network models is significantly more efficient than the others. The gradient boosting method appears to be a little worse but not as much as the linear model which is much less efficient than the other methods: Non-linear approaches improve persistence forecast by up to 34 \% at 3 hour horizon while the linear model improvement only reaches 22 \%. Such a result can be explained intuitively by the fact that a non-linear method can exploit the spatial distribution optimally, as a function of the wind conditions (amplitude and direction) whereas a linear method will necessarily perform an average optimization. Finally, the observed gains at $6$-hour horizon have not significantly increased in comparison to $3$-hour horizon. This is certainly because, in agreement with previous remarks, we don't have considered sufficiently distant sites for such a large time delay.


\begin{figure}[!th]
	\begin{center}
		\hspace*{-0.5cm}
		\includegraphics[width=9cm,angle=0]{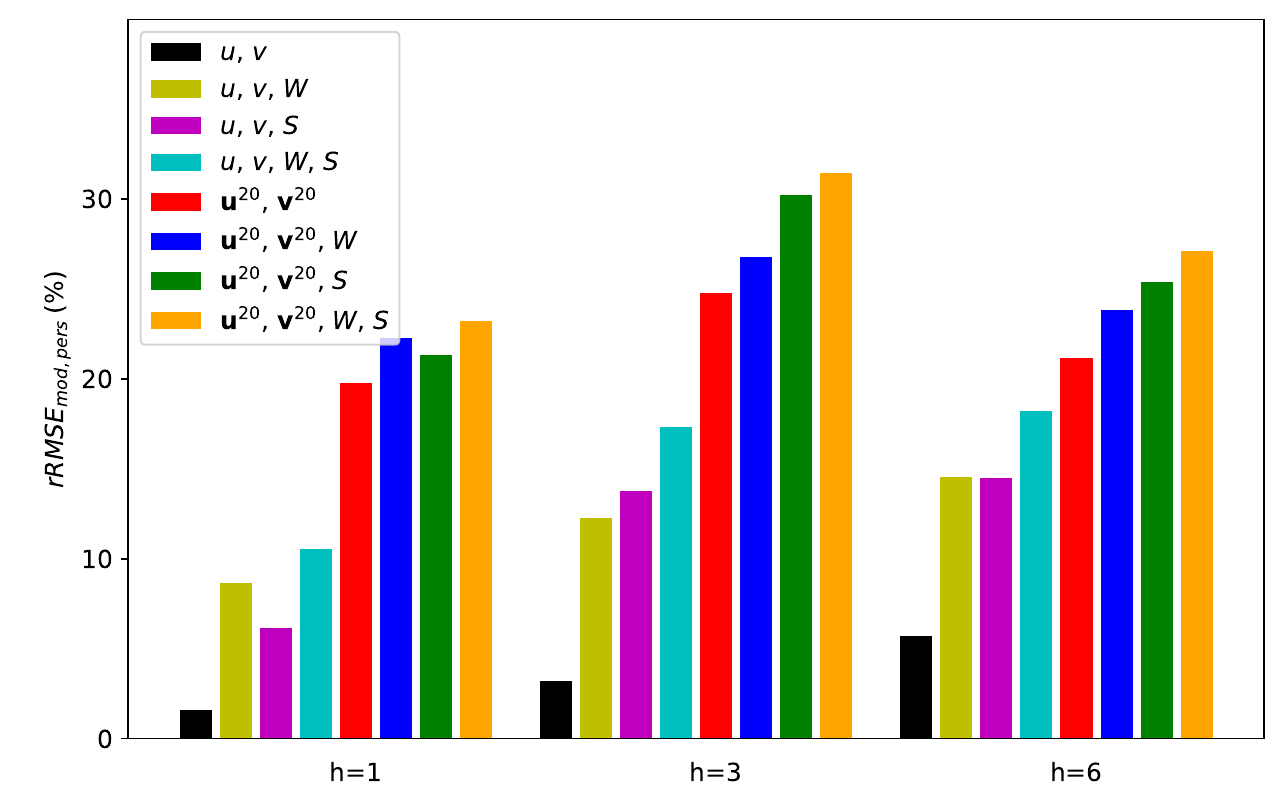}
	\end{center}
	\vspace*{-0.5cm}
	\caption{Average value of the rRMSE values as obtained independently for nine studied sites, with 8 different versions of the fully connected model compared to persistence. The model versions differ by the data used as input : ($u$, $v$) means that the local velocity Cartesian components up to $3 \times h$ hours in the past are considered, ($W$) means that one considers local pressure, temperature and dew-point temperature up to $3 \times h$ hours in the past , ($S$) indicates that one takes into account seasonal/diurnal parameters as encoded by the sine and cosine of the day of the year and the hour within the day and ($\bu^{20}$, $\bv^{20}$) means that the cartesian velocity components of the 20 nearest stations up to 3 hours in the past are used. The higher the bar, the better the forecast as compared to persistence model.}
	\label{fig_input}
\end{figure}

Along the same line, we can also look at the influence of the choice of input variables for a given model on its forecasting performances. Figure \ref{fig_input} illustrates the forecasting accuracy of eight different versions of the fully connected model, compared to persistence (obtained, as before, on average on the 9 sites of the study, for 1, 3 and 6 hours step forward horizon $h$).
We compute the model relative performances when considering as input (i) only the velocity Cartesian components up to $3 \times h$ hours in the past ($u$, $v$), (ii) the velocity components and the local pressure, temperature and dew-point temperature up to $3 \times h$ hours in the past ($u$, $v$, $W$), (iii) the velocity components and the vector of seasonal/diurnal parameters as defined in Eq. \eqref{eq:defS} ($u$, $v$, $S$), (iv) the velocity components, the seasonal components and the weather local variables ($u$, $v$, $W$, $S$), (v) the cartesian velocity components of the 20 nearest stations up to 3 hours in the past ($\bu^{20}$, $\bv^{20}$), (vi) the cartesian velocity components of the 20 nearest stations and the local weather variables ($\bu^{20}$, $\bv^{20}$, $W$), (vii) the cartesian velocity components of the 20 nearest stations and the seasonal/diurnal parameters ($\bu^{20}$, $\bv^{20}$, $S$) and finally, (viii) the cartesian velocity components of the 20 nearest stations, the local weather variables and the seasonal/diurnal components ($\bu^{20}$, $\bv^{20}$, $W$, $S$).
 At horizon 1 and 3 hours, it appears that adding, as input, neighboring site wind data (from the 20 closest stations here) is much more efficient for the quality of the forecast than adding weather data (pressure, temperature, dew point temperature) or seasonal/diurnal data (date and time of the day) ; the gain is less important at a horizon of 6 hours, probably because the neighboring sites are too close to draw the necessary information (see Figs. \ref{fig_addsites},\ref{fig_dist}). It is noteworthy that all the results presented in previous Figs. \ref{fig_addsites}, \ref{fig_dist}, \ref{fig_dir}, \ref{fig_compare} correspond to version (vii), i.e. where only seasonal/diurnal parameters and neighboring stations wind velocities are considered. We can see that adding local weather data only leads to a marginal improvement.

\subsection{Multi-site and multi-step forecasting model}

\begin{figure}[!th]
	\begin{center}
		\hspace*{-0.3cm}
		\includegraphics[width=10cm,angle=0]{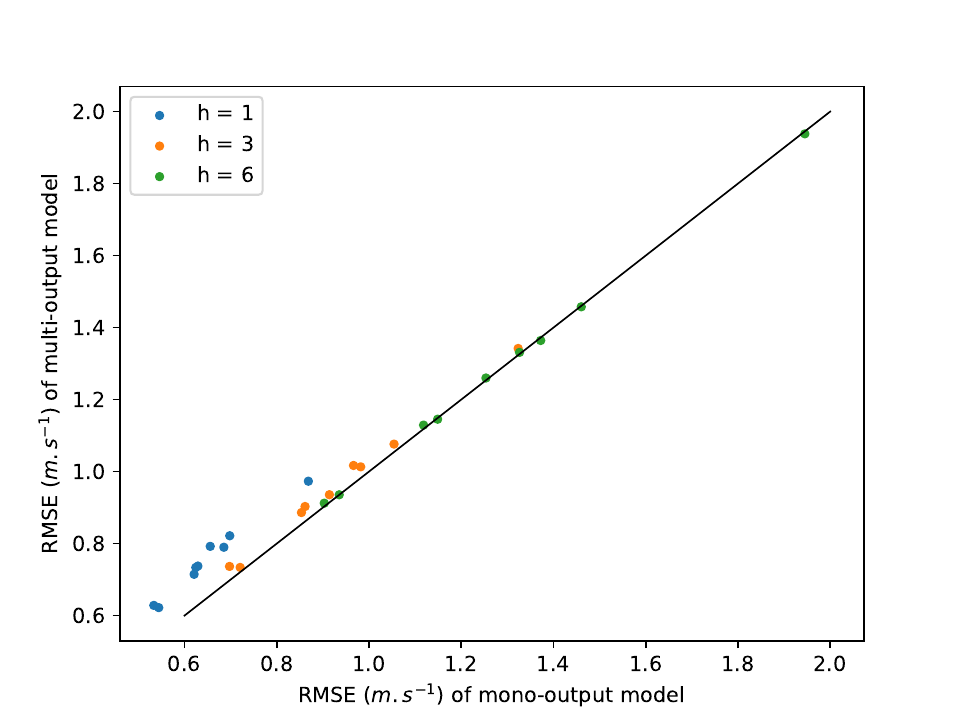}
	\end{center}
	\caption{Comparison of RMSE values obtained for two different model configurations. First, we run the model with only one output : the wind speed at the site of interest (for the 9 sites of the study independently) and at the horizon of interest (here 3 horizons are represented : 1, 3 and 6 hour horizon). Second, we run the model with several outputs : the 9 sites and the 6 horizons are simultaneously outputs of the model, with all the 42 sites as input to the model.}
	\label{fig_allhor}
\end{figure}

As recalled in Section \ref{sec:reg}, machine learning models can handle vector outputs: it is possible to run a model with several outputs, the main interest being to reduce computation time but also to involve fewer parameters and thus to handle a more parsimonious model. Then, we can wonder what the loss of efficiency of the multi-output model is, as compared to the collection of single-output ones; the wind speed data of the 9 studied sites at several horizons are simultaneously predicted (from 1 to 6 hours), with the full 42 station wind data as input to the model. Then the model tries to find the best forecast for all the $O = 54$ outputs (9 sites $\times$ 6 horizons), by minimizing the final MSE. Results are reported in Fig. \ref{fig_allhor}, where RMSE obtained by the multi-output model is represented as a function of RMSE obtained by the single-output models, for 3 forecast horizons (from $1$ hour to $6$ hours). We find that single output models (optimized individually for each site and each horizon) outperform the multi-site multi-horizon model (one model for 54 outputs) for small horizons: for example, the gain is between 12 and 20 \% (depending on the considered site) at $1$-hour horizon and 2 to 9 \% at $2$-hour horizon.
However, we can see that the larger the horizon, the less difference there is between the two approaches and on average, for horizons greater than $3$ hours, the quality of the forecast is equivalent for the two different model configurations.

\section{Conclusion}
\label{s_conc}
In this paper, we have addressed the problem of short-term wind speed forecasting
using different machine learning models: a linear model, a Gradient Boosting Model and four different Deep Neural Network models (a fully connected model, a Graph Convolutional Neural Network, a Recurrent Neural Network model and a Temporal Convolutional model). All these methods have been trained to forecast hourly mean wind speed data up to six hours ahead, in nine different sites distributed in Holland. The obtained results show that all models, with regard to the minimization of the mean-square error, outperform the persistence approach, which represents the simplest reference model.

We have demonstrated that there is a real advantage of using wind speed data from neighboring stations. We have seen that RMSE decreases monotonically when the number of input sites increases and that the higher the forecast horizon, the more interesting it is to add data from geographically distant locations. The direction of the considered neighboring sites is also crucial since it is more interesting to exploit information from stations in the upwind direction. All our results suggest that the forecasting methods are capturing the advection of the surface wind field by itself. Since such a task is by nature non-linear, it is not surprising that non-linear methods provide better results than simple linear regression.  We have finally observed that none of these methods stand out from the rest since they all get the same performance within a few percent differences. 

We also studied the issue related to input variable selection: we have shown that adding wind speed data from neighboring locations as input to the model is more efficient than adding weather (pressure, temperature and dew point temperature) or accounting for seasonal/diurnal variations by considering the date and time of day. Finally, we highlighted the fact that \label{key}a multi-output model that provides a vector output corresponding to the simultaneous forecast at several horizons and for several sites, allows one to handle a more parsimonious model and to considerably reduce computation time without really degrading the forecast at least as far as quite large horizons are concerned.

As a perspective for future research, it would be interesting to know to what extent the ``multi-station" approach is also worth implementing to predict other quantities like the occurrence of rainfall or the level of solar radiation. It would also be interesting to quantitatively analyze how taking into account wind data from very distant sites as input to the model allows one to improve the forecast at high horizons and to compare, at such large horizons, the relative performances of our approach and standard NWP forecasts.  Because of the small geographical extent of the area associated with the KNMI data, we were unable to address this question within the current work. On the opposite, it would also be interesting to exploit better resoluted database in time and space in order to extend our analysis towards smaller time horizons. These last two prospects would allows us to perform a comprehensive study in order to explain the shape of the error curves of Fig. \ref{fig_addsites}. Since one of our main messages is that increasing the number of spatial locations we account for as input is likely to increase a model forecasting performances, in a forthcoming study, we also plan to utilize the whole 2D velocity field as provided by the output of the high-resolution numerical weather prediction model. 

\section*{Acknowledgment}
This work was partially supported by ANR grant SAPHIR project ANR-21-CE04-0014-03.



\end{document}